\documentclass[aip,reprint]{revtex4-1}

\draft 

\usepackage{graphicx}
\usepackage{dcolumn}
\usepackage{bm}
\usepackage{hyperref}

\usepackage[utf8]{inputenc}
\usepackage[T1]{fontenc}
\usepackage{mathptmx}
\usepackage{hyperref}

\usepackage{xcolor}

\begin{document}


\title{Dynamics of fintech terms in news and blogs and specialization of companies of the fintech industry} 



\author{Fabio Ciulla}
\affiliation{Quid, San Francisco, CA 94111, United States}

\author{Rosario N. Mantegna}
\email[]{rosario.mantegna@unipa.it}
\affiliation{Dipartimento di Fisica e Chimica, Universit\`a di Palermo, 90128 Palermo, Italy}
\affiliation{Complexity Science Hub Vienna,1080 Vienna, Austria}
\affiliation{Computer Science Department, University College London, WC1E 6EA London, UK}


\date{\today}

\begin{abstract}
We perform a large scale analysis of a list of fintech terms in (i) news and blogs in English language and (ii) professional descriptions of companies operating in many countries. The occurrence and co-occurrence of fintech terms and locutions shows a progressive evolution of the list of fintech terms in a compact and coherent set of terms used worldwide to describe fintech business activities. By using methods of complex networks that are specifically designed to deal with heterogeneous systems, our analysis of a large set of professional descriptions of companies shows that companies having fintech terms in their description present over-expressions of specific attributes of country, municipality, and economic sector.  By using the approach of statistically validated networks, we detect geographical and economic over-expressions of a set of companies related to the multi-industry, geographically and economically distributed fintech movement.  
\end{abstract}


\maketitle 

\begin{quotation}
We present a study of the rapid development of a highly innovative industry. Specifically, we investigate the fintech industry, i.e. the industry developing technological innovations, technology-based products, and services for the financial sector. This industry presents a rather fast dynamics and a worldwide diffusion. These aspects make very difficult an analysis based on a big data approach due to the unavoidable variety, biases and inconsistencies of the best available databases. In our study, we overcome these limitations by using the methodology of statistically validated networks. In fact, this methodology is  able to highlight over-expressed relationships between pairs of elements of bipartite networks obtained from heterogeneous sets. By investigating a list of terms used in a large corpus of news and blogs and in a large collection of professional descriptions of companies working worldwide, and by using the methodology of statistically validated networks, we detect over-expressions of some fintech terms in the descriptions of companies with specific attributes of geographical location and of economic activity.
\end{quotation}

\section{Introduction}
\label{Int}
Fintech is a term used by several organizations and academics. The term describes research, activities, products, practices, and services  bridging finance, information technology, software development, computer science, and sociology. As for many fruitful and deep concepts the term meaning is not static, fully or uniquely defined \citep{elkana1976discovery} and several attempts have been done to properly frame the concept \citep{schueffel2016taming} and its evolution over time \citep{arner2015evolution}. 
The first written record of ``fintech'' term is found in an academic paper by Abraham Bettinger \citep{bettinger1972fintech}. At that time the term was essentially unnoticed and it was independently reformulated in the early 1990s to describe  a project initiated by Citigroup to facilitate technological cooperation efforts \citep{arner2015evolution}.  The global financial crisis of 2008 and the success of new players delivering financial services by means of technological innovations, particularly in Asia and in emerging countries, has triggered an enormous interest towards fintech challenges and solutions.    

Fintech is today a rapidly growing business area that is active at the interface of many industries all over the world. Tools and services of fintech companies affect (or have a potential to affect) many traditional and new areas of finance. The impact of fintech companies also extends well beyond the field of finance. Examples are products and services such as the ones associated with the use of the blockchain in the food supply chain or in the monitoring of infectious diseases.

In this contribution, we aim at answering two scientific questions. The first question asks whether some terms referred to products, services, and methods are jointly used to describe fintech activities in news and blogs in recent years. We answer this question by investigating a large corpus of texts of news and blogs sources written in English collected over the Internet during the years from 2014 to 2018. The corpus is investigated with basic tools of network science  \citep{borner2007network,newman2010networks,barabasi2016network}. Specifically, starting from a list of terms (composed of single or multiple words) highlighted by experts, we investigated the network of co-occurrence of pairs of terms in a large corpus of texts of news and blogs for each calendar year of the database. We verify that the network of co-occurrence becomes progressively more dense and topologically compact supporting the hypothesis that this group of terms describes business and technological activities addressed by the general term fintech. 

The second scientific question focuses on the profile of companies with fintech interests or activities operating in many countries. Specifically, we investigate economic sector, country, and municipality of a very large number of companies located worldwide by using the list of terms selected in the first part of our study and by detecting their presence in the descriptions of companies that are present in the professional databases Capital IQ and Crunchbase. 
We show that the over-expression of economic sector, country (more precisely country or dependent territory), and municipality of the headquarter of the company presents two statistical regularities: (i) some companies dealing with fintech processes, products or methods specialize on specific fintech sub-topics; (ii) some companies concentrate their activities in specific economic sectors and/or in specific geographical clusters.

This second investigation presents an important challenge due to the fact that the coverage of the databases is geographically heterogeneous with a special focus on western countries. To overcome this problem of bias of databases towards western countries, we leverage on a methodology developed in network science  \citep{serrano2009extracting,tumminello2011statistically}. This methodology is based on the study of statistically validated networks \citep{tumminello2011statistically,hatzopoulos2015quantifying}, and it is able to detect over-expressions of linkages in heterogeneous networks successfully overcoming the problem of the heterogeneity and bias of the coverage of databases.

By applying the methodology of statistically validated networks, we first construct three bipartite networks and we then analyze them to detect over-expressions of linkages that are present between (i) economic sectors, (ii) countries, and (iii) municipalities of companies and fintech terms characterizing different areas of fintech products, services and activities such as, for example, {\it financial inclusion}, {\it anti-money laundering}, etc. In other words, our methodology highlights specializations of sets of companies in an heterogenous setting allowing to obtain statistically significant results starting from a heterogeneous source of data.

The paper is organized as follows. In section \ref{Fwd} we describe a set of selected fintech terms and the investigated databases. Section \ref{Res} presents the empirical results obtained in the analysis of networks of co-occurrence of fintech terms sampled at different calendar years. In section \ref{Apd} we investigate over-expressions detected in the bipartite networks of (i) economic sectors and fintech terms, (ii) countries and fintech terms, and (iii) municipalities and fintech terms. Section \ref{Dco} discusses the results obtained and presents some conclusions.  

\section{Fintech terms and datasets}
\label{Fwd}
In this paper, we investigate the occurrence and co-occurrence of a set of 53 fintech terms.  The set is selected starting from the analysis of a series of fintech terms collected and commented by experts in several web pages. One example of these lists of terms can be  accessed at the web page reporting the article "Fintech lingo explained" by Anna Irrera and Maria Caspani
{\it https://www.reuters.com/article/us-usa-fintech-explainer-idUSKBN19D29I}  \citep{accessweb}. Other examples of web pages with fintech list of terms are:
(i) {\it https://eba.europa.eu/financial-innovation-and-fintech/glossary-for-financial-innovation}, 
(ii) {\it https://www.nbs.sk/en/financial-market-supervision1/fintech/fintech-glossary},
(iii) {\it https://www2.deloitte.com/uk/en/pages/financial-services/articles/fintech-glossary.html}.

The 53 investigated terms are listed in Table~\ref{tabwords}. They include (a) words like bitcoin, blockchain, crowdfunding, (b) groups of words expressing a precise concept such as anti-money laundering, combating the financing of terrorism, etc. , (c) word contractions such as fintech, finserv, and segwit (together with their expanded terms), and (d) acronyms (saas and EMV). It is worth stressing that we have used acronyms only in the absence of polysemy. For example, we did not use the widely used acronym AML for anti-money laundering because it is also frequently used for acute myeloid leukemia, which is a distinct concept.    

\begin{table*}
\scriptsize
\begin{tabular}{lll}  
\hline
anti-money laundering & genesis block & robo-advisors \\
bitcoin & hard fork & (automate investment advice) \\
blockchain & hash rate & saas \\
card not present & high speed networks & (software-as-a-service) \\
chief data officer & initial coin offering & segwit \\
collaborative consumption & insurtech & (segregated witness) \\
collaborative economy & know your customer & sharding \\
combating the financing of terrorism & knowledge-based authentication & single sign-on authentication \\
counter-terrorist financing & messaging commerce & smart contracts \\
crowdfunding & on-boarding & (blockchain-based contracts) \\
cryptocurrency & open banking & social lending\\
digital wallet & p2p lending &  soft fork \\
distributed ledger technology & (peer-to-peer lending) & sybil attack \\
emv chip & payment gateway & token sale \\
(Europay, MasterCard, and Visa) & pci compliance & tokenization \\
equity-crowdfunding & (payment card industry compliance) & unbanked \\
ethereum blockchain & point-of-sale & underbanked \\
financial inclusion & proof-of-authority & user as owner \\
finserv & proof-of-stake & virtual currency \\
(financial services industry) & proof-of-work & ~ \\
fintech & regtech & ~ \\
(financial technology) & (regulatory technology) & ~ \\
 \hline
\end{tabular}
\caption{List of fintech terms investigated in our study. Terms are listed in alphabetical order from the first to the third column. The terms in parenthesis are expanded variants of the previous term.}
\label{tabwords}
\end{table*}

Our first investigation concerns the occurrence and co-occurrence of fintech terms in texts of a corpus of news and blogs. The database of news and blogs covers texts distributed over the Internet during the calendar years of 2014, 2015, 2016, 2017, and 2018. It consists of approximately 1 billion texts written in English language collected by considering approximately 60,000 news sources and 500,000 blogs. The corpus is a proprietary corpus of the company LexisNexis. The geographical origin of text sources is primarily located in the United States (47.5\% of texts) and in the United Kingdom (15.4\% of texts). The remaining 37.1\% of texts originates from 207 different sovereign countries or overseas territories or dependent territories or unincorporated territories such as, for example, Hong Kong, Macau, Greenland, Puerto Rico, Faroe islands, Falkland islands, etc. . For the sake of simplicity, in the following paragraphs we use the word country to describe an entity being  a sovereign country or an overseas territory or a dependent territory or an unincorporated territory or a similar type of institution. In this corpus, we investigate the occurrence and co-occurrence of fintech terms to track the evolution of the use of our selected terms of fintech products and services in the English language in recent years. 

In our second investigation, the occurrence of selected fintech terms is investigated in the professional description of companies operating in many countries. The dataset of company descriptions is a dataset curated by the Quid company. The dataset was obtained by merging the information present in two proprietary databases. These databases are the Capital IQ database of S\&P Global company and the Crunchbase Pro database of Crunchbase company. Capital IQ database provides a quite complete coverage of publicly-listed companies. In fact, the database covers 99\% of global market capitalization  according to Capital IQ website. Crunchbase database is more focused on innovative companies although currently also covers public and private companies on a global scale. 
Our dataset is obtained  from the merging and pre-processing of the two databases. 
The total number of company descriptions is about 2.2 million. They are descriptions of companies with headquarters located in 239 different countries (where country has the broadly defined meaning clarified above), and classified as working in 68 different economic sectors. Although the dataset covers a large part of global market capitalization, it is not unbiased. In fact, a very high percent of companies are located in United States (61.3\%) and in United Kingdom (7.50\%) indicating that most small and innovative companies included in the datasets are operating in these two countries. Other top represented countries are China (2.48\%), Germany (1.99\%), France (1.76\%), India (1.60\%), Canada (1.51\%), Italy (1.38\%), Spain (1.35\%), and Australia (1.28\%). The bias is reduced but still present when we only consider public companies. For public companies the ten top countries with highest percent of companies are United States (29.3\%), Canada (10.3\%), China (7.36\%), India (6.32\%), Japan (5.50\%), United Kingdom (3.72\%), Australia (3.51\%), South Korea (3.25\%), Taiwan (2.59\%), and Hong Kong (2.37\%). In our analysis, we therefore need to take into account the bias that is present in the dataset. In Section \ref{Apd} we will take into account the bias by using a statistical methodology of network science that is able to highlight over-expression in bipartite networks in the presence of a pronounced heterogeneity of the elements (in the present case the attributes of companies). Both texts of news and blogs, as well as texts of companies' descriptions, have been indexed and queried using the open-core Elasticsearch search engine.

\begin{table*}
\scriptsize
\centering
\begin{tabular}{|lrrrrrr|r|}  
\hline
Fintech term  & News & News & News & News & News & News & Companies \\
~  & and blogs & and blogs & and blogs & and blogs & and blogs & and blogs & descriptions \\
~  & 2014 & 2015 & 2016 & 2017 & 2018 & all years & ~ \\
\hline
software as a service (saas) & 669549 & 745176 & 559525 & 482543 & 509814 & 2966607	&	14210 \\
bitcoin & 196728 & 158893 & 127020 & 385084 & 1595799 & 2463524	&	1785\\
cryptocurrency & 31182 & 33573 & 31566 & 207403 & 1671363 &	1975087	&	1908\\
blockchain & 11391 & 46935 & 118145 & 371307 & 1009427 &	1557205	&	6378\\
fintech & 89435 & 197436 & 321873 & 421670 & 498991 &	1529405	&	5331\\
crowdfunding & 201681 & 288203 & 253103 & 223953 & 222131 &	1189071	&	1996\\
point-of-sale & 267858 & 275910 & 209134 &186231 & 203124 &	1142257	&	5230\\
finserv & 187031 & 224312 & 195813 & 180241 & 154649 &	942046	&	1224\\
anti-money laundering & 46586 & 60800 & 73999 & 76564 & 96464 &	354413	&	359\\
financial inclusion & 38048 & 54993 & 69253 & 73368 & 86089 &	321751	&	268\\
virtual currency & 52121 & 31796 & 29339 & 54565 & 70715 &	238536	&	246\\
on-boarding & 35901 & 44238 & 40336 & 38952 & 35782 &	195209	&	459\\
proof-of-work & 1152 & 1889 & 1893 & 4235 & 180364 &	189533	&	32\\
smart contracts & 523 & 3221 & 12160 & 39983 & 105688 &	161575	&	521\\
unbanked	& 27147 &	 30378 & 29342 & 32052 & 39973 &	158892	&	222\\
payment gateway & 30805 & 36781 & 40530 & 20857 & 26558 &	155531	&	765\\
digital wallet & 29101 & 22795 & 21976 & 24242 & 30001 &	128115	&	194\\
tokenization & 21083	 & 34056 & 18966 & 	20855 & 29927 &	124887	&	173\\
know your customer & 15455 & 18448 & 19941 & 24062 & 34547 &	112453	&	135\\
p2p lending &15812 & 24963 & 30043 & 18377 & 19765 &	108960	&	382\\
proof-of-stake & 828 & 1078 & 1465 & 3656 & 97793 &	104820	&	34\\
emv chip & 18534 & 31545 & 22306 & 10731 & 10650 &	93766	&	39\\
pci compliance& 25918 & 27098 & 11582 & 8542 & 8129 &	81269	&	194\\
distributed ledger technology & 20 & 2122 &10954 & 22064 & 44991 &	80151	&	147\\
initial coin offering &	256 & 3 &	1100 & 23440 & 46168 &	70967	&	63\\
equity-crowdfunding & 9907 & 19297 & 16938 & 14062 & 9771 &	69975	&	201\\
insurtech & 19 & 31 & 6071 & 30857 & 31145 &	68123	&	269\\
ethereum blockchain & 7 & 362 & 2701 & 16925 & 46340 &	66335	&	168\\
underbanked & 10165 & 11953 & 11749 & 10525 & 18639 &	63031	&	109\\
token sale & 8 & 212 & 79 & 23079 & 32848 &	56226	&	47\\
card not present & 13944 & 15682 &	10721 & 5844 & 6079 &	52270	&	87 \\
robo-advisors &	 2719 & 7253 &	18315 & 10885 & 8299 &	47471	&	21\\
regtech & 1455 & 4153 & 6233 & 16116 & 19139 &	47096	&	137\\
chief data officer & 4339 & 9167 & 9038 & 8217 & 11470 &	42231	&	2\\
open banking & 282 & 671 & 2733 & 11227 & 23122 &	38035	&	47\\
high speed networks & 5547 & 6328 & 4233 & 4403 & 4227 &	24738	&	37\\
hard fork & 22 & 148 & 709 & 6013 & 17161 &	24053	&	2\\
collaborative economy & 2125 & 4575 & 2914 & 1851 & 1537 &	13002	&	47\\
collaborative consumption & 4935 & 3694 & 1978 & 820 & 721 &	12148	&	83\\
sharding & 2949 & 2301 & 1258 & 1631 & 3823 &	11962	&	17\\
counter-terrorist financing & 946 & 1070 & 2498 & 2542 & 3170 &	10226	&	9\\
segwit & 5 & 22 & 260 & 3825 & 5129 &	9241	&	2\\
hash rate & 896 & 461 & 275 & 1201 & 5605 &	8438	&	4\\
combating the financing of terrorism & 1072 & 790 &1756 & 2012 & 2518 &	8148	&	2\\
knowledge-based authentication & 1828 & 726 & 1089 & 976 & 1402 &	6021	&	11\\
single sign-on authentication & 1694 & 1593 & 669 & 770 & 461 &	5187	&	6 \\
genesis block & 381 & 55 & 309 & 495 & 3938 &	5178	&	4\\
social lending & 608 & 599 & 829 & 1011 & 720 &	3767	&	31\\
proof-of-authority & 30 & 55 & 193 & 272 & 1407 &	1957	&	2\\
soft fork & 7 & 10 & 210 & 688 & 910 &	1825	&	1\\
messaging commerce & 26 & 9 & 43 & 327 & 49 & 454 & 0\\
sybil attack & 107 & 18 & 45 & 24 & 197 & 391 & 0\\
user as owner & 1 & 3 & 1 & 0 & 0 & 5 & 0\\
\hline
Total occurrences of fintech terms & 2,080,169 & 2,487,880 &	2,355,211 & 3,131,576 & 7,088,729 & ~ & 43,641 \\
Texts with at least one fintech term& 1,418,726 & 1,690,290 & 1,589,152 & 1,887,749 & 3,742,348 & ~ & 38,648 \\
Number of texts in the corpus & 136,048,047 & 172,912,445 &	182,959,692 & 169,448,198 & 175,559,955 & ~ & $2.2 x 10^6$ \\

\hline
\end{tabular}
\caption{Occurrence  of fintech terms in the texts of corpus of news and blogs for each investigated calendar year (from second to sixth column). Occurrence is ranked from top to bottom with the rank of the term defined by the total number of occurrences observed in the 5 years (seventh column). The last column, labeled as ``Companies descriptions", shows the occurrence of the fintech terms in the professional documents describing companies included into Capital IQ and Crunchbase Pro databases.}
\label{tabcounts}
\end{table*}

\begin{table}
\centering
\begin{tabular}{|r|rrrrrr|}  
\hline
Year  & \# nodes & \# edges & edge & \# connected & Average & Diameter \\
~  & ~ & ~ & density & components & path length & ~ \\

\hline
2014 & 46 & 483 & 0.467 & 1 & 1.54 & 3  \\
2015 & 51 & 625 & 0.490 & 1 & 1.52 & 3  \\
2016 & 52 & 823 & 0.621 & 1 & 1.38 & 3  \\
2017 & 53 & 950 & 0.689 & 1 & 1.31 & 2  \\
2018 & 52 & 1002 & 0.756 & 1 & 1.24 & 2  \\
\hline
\end{tabular}
\caption{Number of nodes, number of edges, edge density, number of connected components, average path length, and diameter of fintech term co-occurrence networks for each calendar year. The co-occurrence network of fintech terms is obtained by analyzing texts of a corpus of approximately 1 billion texts collected from news and blogs.}
\label{tabdensity}
\end{table}

\section{Results on the analysis of texts of news and blogs}
\label{Res}
We first search the fintech terms in the texts of the corpus of news and blogs for the calendar years from 2014 to 2018. The counts obtained are shown in Table~\ref{tabcounts}. The Table shows that the occurrence of the 53 fintech terms is quite heterogeneous ranging from the 1,671,363 occurrences of  {\it cryptocurrency} in 2018 to no occurrence of {\it user as owner} in 2017 and 2018.  The pronounced heterogeneity is not too surprising due to the fact that the fintech list of terms comprises both quite wide concepts such as, for example, {\it software as a service} and very specialized concepts such as, for example, {\it hard fork} or {\it soft fork}. The number of texts investigated changes only moderately over the years. Their values are reported in the last row of Table~\ref{tabcounts}. The minimum number of texts investigated in a year was about 136 millions in 2014 and the maximum was about 183 millions in 2016. The average value was 167 millions with a standard deviation of 18.2 millions, i.e. only about 11 percent of the average value. In the bottom part of Table~\ref{tabcounts}, we also show the total occurrence of fintech terms per year and the number of texts with at least one fintech term.
 
 \begin{figure}[h!]
\begin{center}
\includegraphics[width=1.1\linewidth]{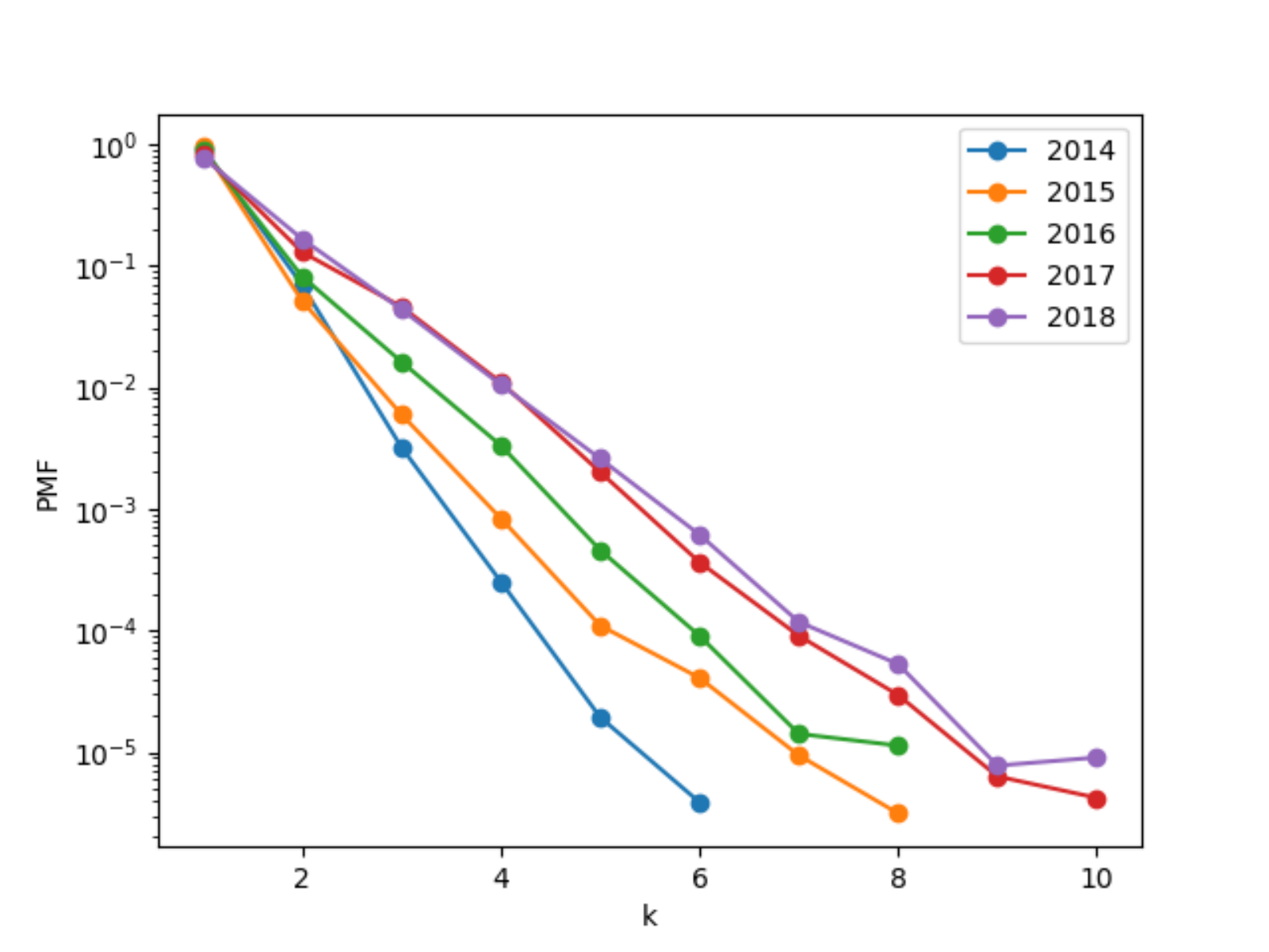}
\end{center}
\caption{Probability mass function of the number of co-occurrences of fintech terms detected in a single text. Symbols of different colors refer to different years. The vertical axis is logarithmic.}
\label{fig:dist}
\end{figure}

For some terms, we note a quite pronounced variation of the occurrence. For example, {\t bitcoin, cryptocurrency, blockchain, smart contracts,  insurtech and regtech} show prominent large variations of the occurrences in a relatively limited period of time. The occurrence analysis is therefore highlighting heterogeneity of the fintech terms and also a pronounced dynamics of some of them. We interpret this dynamics as an indication of the process of definition and specialization of the new terms. Let us consider, for example, the two terms fintech and finserv. These two terms are connoting different aspects of technological applications and service solutions of specific financial problems. The semantic difference between the two terms is debated over the years (see for example the 2015 blog https://finiculture.com/finserv-fintech/ for an opinion about it). The occurrence dynamics of the two terms observed from 2014 to 2018 shows a clear pattern. The term finserv has a pattern of decreasing occurrence while the reverse is true for fintech. In other words although in past years the two terms have been both used with a similar level of diffusion, in most recent years  fintech is emerging as the term describing both technological solutions and digital services applied to financial innovations.


\begin{figure*}[th!]
\begin{center}
\includegraphics[width=1.0\linewidth]{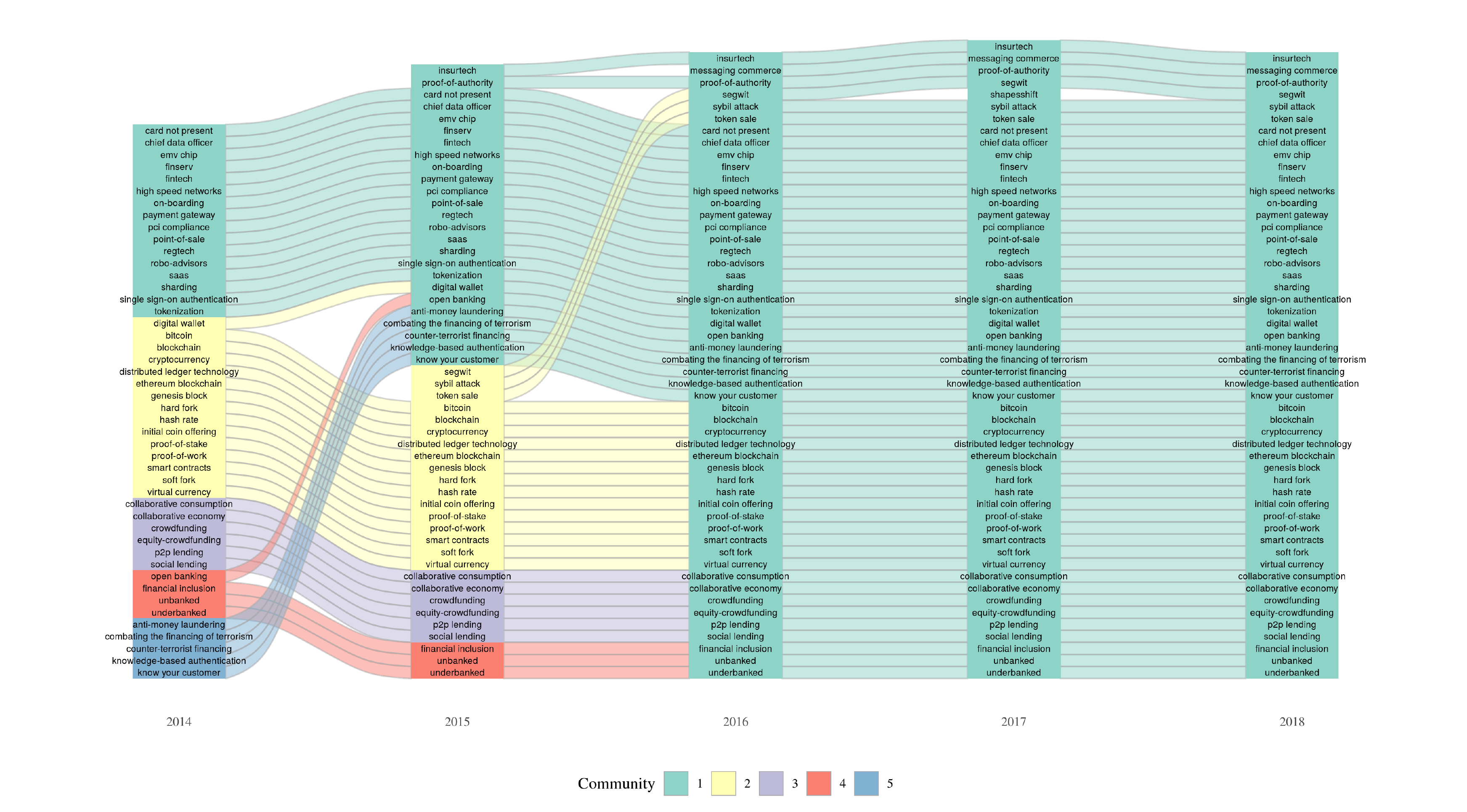}
\end{center}
\caption{Alluvial diagram of the communities found by the Infomap algorithm on the weighted co-occurrence networks of the years from 2014 (left) to 2018 (right). Each vertical set refers to a year. Colors are defining different communities. Fintech terms are shown in each box. We detect five communities in 2014, four communities in 2015 and one community from 2016 to 2018.}
\label{fig:Alluvial}
\end{figure*}


The second type of investigation concerns the co-occurrence of pairs of fintech terms in the same text. In this investigation, we start to make use of networks as an analysis tool, indeed fintech terms are represented as nodes and an edge exists between two nodes when the two fintech terms are present in the same text at least once. In Table \ref{tabdensity} we show the time evolution of the number of nodes and edges of the network of co-occurrence of fintech terms. The Table shows that the co-occurrence network is always characterized by a number of nodes very close to the number of investigated terms and by a number of edges that is growing from 2014 to 2018. 
In all years, we detect a single connected component and the network edge density  is growing from 0.467 (in 2014) to 0.756 (in 2018). In parallel with the edge density increases we also detect a steadily decreases of the average path length. The diameter of the network, i.e. the longest distance between any two terms in number of steps, is 3 for the 2014-2016 years and jumps to 2 in the last two years. The network is therefore highly dense and compact in the investigated years.

By performing numerical simulations, we have verified that the topology of the unweighted co-occurrence network is consistent with the one of an Erd\"os-R\'enyi model \citep{newman2010networks,barabasi2016network} with the same number of nodes and edges. However, the consistency of the empirical topology with an Erd\"os-R\'enyi topology does not mean that the co-occurrence of words is a random phenomenon. In fact, hereafter we show that  a null hypothesis of random matching of two different terms in the same text is not consistent with the observed value  $N_{A,B}$ of co-occurrence of terms $A$ and $B$. In our null model,  the probability of occurrence of each term $A$ is $P(A)$. By assuming a completely random matching of two terms $A$ and $B$ in the same text, the probability of observing a co-occurrence is the product of $P(A)$ times $P(B)$. Starting from this probability and assuming as a null model a binomial distribution with probability $P(A) P(B)$, the expected value $E[N_{A,B}]$ of the co-occurrence is given by $N_T P(A) P(B)$ where $N_T$ is the total number of texts analyzed. The standard deviation of the same variable is $\sqrt{N_T P(A) P(B)(1-P(A) P(B))}$. Under this null hypothesis, for each pair of terms, we estimate a  z-score by computing

\begin{equation}
\label{pvaltheor}
z(A,B)=\frac{N_{A,B}-E[N_{A,B}]}{SD[N_{A,B}]}=\frac{N_{A,B}-N_T P(A) P(B)}{\sqrt{N_T P(A) P(B)(1-P(A) P(B))}}.
\end{equation}
By analyzing the $z$-score values for all pairs of terms of the co-occurrence networks we verify that $z$ values are very large and in all cases they exceed 3 for a fraction of edges ranging 
from 80.0 percent (in 2014) to 91.3 percent (in 2017). In summary, almost all detected co-occurrences of pairs of terms are not consistent with a random matching of the terms and they suggest that their joint use carry information in the text.

We also verify that the detected co-occurrences are not originated in a limited number of texts including the presence of many of the terms investigated. In Fig.~\ref{fig:dist} we show the probability mass function of observing $k$ co-occurrences of fintech terms in a single text. The probability mass function is shown in a semi-logarithmic plot and it is well approximated by an exponentially decaying function. The figure shows that  multiple co-occurrences increases in texts from 2014 to 2018 but the largest majority of texts presents just a single co-occurrence of fintech terms.  

To further verify the role of the heterogeneity of the number of co-occurrences, we characterize the co-occurrence network as a weighted network where the weight of a link between node $A$ and node $B$ is given by the co-occurrence $N_{A,B}$. In this weighted network, we perform a community detection analysis with the algorithm Infomap \cite{rosvall2008maps} to search for any internal structure of the co-occurrence networks. The Infomap algorithm is one of the most widely used community detection algorithms. It can be applied both to unweighted and weighted networks. We apply the Infomap algorithm to the weighted co-occurrence networks and we find the communities shown in Fig.~\ref{fig:Alluvial}. The algorithm detects five communities in 2014, four communities in 2015 and a single community starting from 2016. In summary, the weighted co-occurrence networks are becoming denser over time. We interpret the time evolution of the weighted co-occurrence network as the progressive setting of a coherent set of terms used in the business and technology area generically addressed as fintech.  In the next Section, we will use this set of fintech terms to investigate the professional descriptions (written in English language) of a large and heterogeneous set of companies operating in many countries.

\section{Analysis of professional descriptions of companies}
\label{Apd}

In this section, we report on the analysis of fintech term occurrences detected in professional documents (i.e. documents written by economic analysts)  describing the profile of companies operating in many countries. These are the descriptions of companies that are present in the Capital IQ database and in the Crunchbase Pro database. This set of professional texts is a relatively limited corpus comprising 2.2 million documents. 

We detect at least one term of the fintech list in 38,648 distinct descriptions of companies. We believe this number can be considered as a rough estimation of the number of companies currently focused on fintech. In fact the number is about 3 times the estimate made by a McKinsey study in 2016 \citep{drummer2016fintech}. In the last column of Table~\ref{tabcounts}, we report the occurrence of the 53 fintech terms in the documents of the dataset. Specifically, 50 out of 53 terms are detected in the documents describing the companies. The occurrence profile of the terms is pretty similar to the occurrence profile detected in the corpus of texts of news and blogs. In fact the correlation between the occurrence of the 50 terms detected both in the texts of news and blogs and in the descriptions of companies is 0.824 (when measured as the Pearson's correlation coefficient between term occurrence) or 0.891 (when measured as the Spearman's correlation coefficient between term rank). 
This similarity of use of fintech terms in news and blogs and in professionally edited texts is another evidence supporting the assumption that the set of fintech terms defines a compact and coherent set of terms.

\begin{table*}
\scriptsize
\centering
\begin{tabular}{||l|r||l|r||l|r||}  
\hline
Country  & Occurrence & Industry & occurrence & Municipality & Occurrence \\
\hline
United States	&	15502	&	Internet Software and Services	&	13891	&	London	&	1720	\\
United Kingdom	&	2934	&	Software	&	8582	&	New York	&	1566	\\
Canada	&	1847	&	Capital Markets	&	3729	&	San Francisco	&	1216	\\
China	&	1317	&	IT Services	&	2899	&	Singapore	&	669	\\
India	&	1237	&	Media	&	896	&	Paris	&	470	\\
Germany	&	964	&	Professional Services	&	893	&	Toronto	&	457	\\
France	&	907	&	Health Care Technology	&	646	&	Beijing	&	436	\\
Australia	&	772	&	Electronic Equipment, Instruments and Components	&	559	&	Chicago	&	401	\\
Singapore	&	680	&	Commercial Services and Supplies	&	502	&	Los Angeles	&	318	\\
Switzerland	&	457	&	Diversified Financial Services	&	466	&	Boston	&	316	\\
Israel	&	451	&	Banks	&	426	&	Berlin	&	307	\\
Spain	&	434	&	Consumer Finance	&	364	&	Vancouver	&	291	\\
Brazil	&	432	&	Technology Hardware, Storage and Peripherals	&	223	&	Austin	&	283	\\
Netherlands	&	416	&	Insurance	&	149	&	Atlanta	&	279	\\
Hong Kong	&	346	&	Real Estate Management and Development	&	126	&	Shanghai	&	279	\\
Japan	&	304	&	Hotels, Restaurants and Leisure	&	124	&	Palo Alto	&	267	\\
Ireland	&	291	&	Diversified Consumer Services	&	122	&	Mumbai	&	241	\\
Italy	&	256	&	Diversified Telecommunication Services	&	106	&	Tokyo	&	231	\\
Sweden	&	237	&	Internet and Direct Marketing Retail	&	95	&	Sydney	&	225	\\
South Africa	&	229	&	Communications Equipment	&	91	&	Seattle	&	223	\\
Russia	&	224	&	Containers and Packaging	&	81	&	San Diego	&	210	\\
Finland	&	179	&	Healthcare Providers and Services	&	73	&	Dublin	&	205	\\
Poland	&	175	&	Metals and Mining	&	68	&	Tel Aviv	&	181	\\
South Korea	&	170	&	Distributors	&	47	&	Dallas	&	169	\\
Denmark	&	159	&	Machinery	&	41	&	Amsterdam	&	168	\\
Belgium	&	152	&	Trading Companies and Distributors	&	39	&	Denver	&	165	\\
Mexico	&	145	&	Semiconductors and Semiconductor Equipment	&	34	&	Washington	&	159	\\
New Zealand	&	144	&	Air Freight and Logistics	&	33	&	Melbourne	&	154	\\
United Arab Emirates	&	127	&	Construction and Engineering	&	33	&	Miami	&	154	\\
Austria	&	119	&	Wireless Telecommunication Services	&	33	&	Stockholm	&	153	\\
Malaysia	&	118	&	Chemicals	&	31	&	San Jose	&	152	\\
Estonia	&	117	&	Household Durables	&	31	&	Barcelona	&	151	\\
Norway	&	106	&	Specialty Retail	&	29	&	Hong Kong	&	150	\\
Indonesia	&	104	&	Thrifts and Mortgage Finance	&	27	&	Moscow	&	145	\\
Argentina	&	101	&	Textiles, Apparel and Luxury Goods	&	26	&	Shenzhen	&	145	\\
Nigeria	&	91	&	Electrical Equipment	&	25	&	Madrid	&	142	\\
Turkey	&	87	&	Food Products	&	25	&	Mountain View	&	133	\\
Philippines	&	84	&	Industrial Conglomerates	&	24	&	Menlo Park	&	132	\\
Taiwan	&	82	&	Paper and Forest Products	&	22	&	Bangalore	&	130	\\
Ukraine	&	79	&	Road and Rail	&	19	&	Seoul	&	128	\\
Chile	&	73	&	Healthcare Equipment and Supplies	&	18	&	Munich	&	127	\\
Portugal	&	70	&	Aerospace and Defense	&	16	&	Houston	&	122	\\
Luxembourg	&	69	&	Biotechnology	&	14	&	San Mateo	&	121	\\
Thailand	&	63	&	Beverages	&	12	&	Sao Paulo	&	119	\\
Czech Republic	&	61	&	Food and Staples Retailing	&	12	&	Zug	&	116	\\
Malta	&	56	&	Independent Power and Renewable Electricity Producers	&	10	&	Las Vegas	&	115\\
Lithuania	&	55	&	Life Sciences Tools and Services	&	10	&	Cambridge	&	113	\\
Bulgaria	&	54	&	Oil, Gas and Consumable Fuels	&	10	&	Dubai	&	110	\\
Cayman Islands	&	54	&	Airlines	&	6	&	Sunnyvale	 &	110	\\
Vietnam	&	50	&	Personal Products	&	6	&	Irvine &	108	\\
 $\dots$	&	$\dots$	&	$\dots$	&	$\dots$	&	$\dots$	&	$\dots$	\\
$\dots$	&	$\dots$	&	$\dots$	&	$\dots$	&	$\dots$	&	$\dots$	\\
\hline
Unknown	&	4479	&	Unknown	&	2867	&	Unknown	&	5280	\\
\hline
\end{tabular}
\caption{Occurrence of the top 50 most common attributes of country (first and second column), economic sector (third and fourth column), and municipality (fifth and sixth column) of the companies presenting at least one fintech term in their company description. We also provide the total number of unknown for each type of attribute. Companies with at least one fintech term in their description belong to 163 countries, 63 industries and 4474 municipalities.}
\label{tabOrg}
\end{table*}

The databases have a number of attributes characterizing the companies. In the present study, we select country, municipality of the headquarter, and economic sector among them.  A partial summary of these attributes is shown in Table~\ref{tabOrg}. The Table shows the 50 most common attributes of country (first and second column), economic sector (third and fourth column), and municipality (fifth and sixth column) with their occurrence. The Table shows that the occurrence of all three attributes is heterogeneous. To provide a measure of the heterogeneity of occurrences we use the Herfindal index \citep{rhoades1993herfindahl} that is a widespread simple measure of concentration of attributes of a set of elements.  The Herfindhal index $H$ of the reported attributes is  $H=0.223$ for countries, $H=0.228$ for economic sectors, and $H=0.0117$ for municipalities. High values of Herfindhal index indicate high concentration of the attribute in few elements whereas low values indicate homogeneous distribution of the attribute to the different elements. The maximum value of the Herfindal index is one (complete concentration in one element). The minimum value of the Herfindhal index is equal to $H_{min}=1/N_e$ where $N_e$ is the number of elements. In the present case the minimum value ( perfect homogeneity) would be observed when $H_{min}=0.00613$ for countries, $H_{min}=0.0159$ for economic sectors, and $H_{min}=0.000218$ for municipalities. The empirically observed values are all much above the values expected for homogeneous distributions of the attributes and indicate a high degree of heterogeneity. The heterogeneity of attributes reflects both the different diffusion of fintech interest and activities in different countries, municipalities and economic sectors and the heterogeneity of the databases discussed in Section \ref{Fwd}.

The bias of the databases and the heterogeneity of attributes make frequency analysis of the attributes not reliable. We therefore perform an over-expression analysis of the attributes observed in our datasets with a methodology used in network science. With this approach, we highlight over-expression of the presence of some fintech terms in the description  of companies with different attributes of economic sector, country, and municipality of headquarters. This is achieved by selecting those pairwise relationships between an attribute of companies and fintech terms that cannot be explained by a null model of random connection that takes into account the heterogeneity of the attribute and of the fintech terms.

Let us comment in some detail the heterogeneity of the three investigated attributes. The country with the highest number of companies having fintech terms in their professional description is United States. This is consistent both with the bias of databases (in the original set 61.3\% of companies are located in this country) and with the leading role that this country has in the fintech movement. However, in the set of companies having at least one fintech term in their description United States has  40.1\% of companies. This percent is still very high but less than the one observed in the original dataset.  United Kingdom has almost the same percent in the original (7.50\%) and in the selected set (7.59\%). A number of countries that we could label as innovative have higher percent in the selected set. For example, Canada has 1.51\% in the original set and 4.78\% in the selected set. Singapore has 0.378 \% in the original set and 1.76\% in the selected set. Israel has 0.244\% in the original set and 1.17\% in the selected set. Switzerland has 0.967\% in the original set and  1.18\% in the selected set.  We interpret this change of the ranking as an indication that the databases are moderately less biased towards United States and United Kingdom when the coverage focus on companies dealing with fintech topics, methods or products. However, the bias is still quite strong and our analysis will explicitly take into account this limitation of the databases.

To characterize the economic sector, we use the industry classification of the Global Industry Classification Standard (GICS) developed jointly by Standard \& Poor's and MSCI/Barra companies. GICS  was developed in 1999 and it is periodically updated. The GICS structure today is organized in 11 sectors, 24 industry groups, 68 industries and 158 sub-industries. In our analysis we use the classification at the level of industries of July 2018. The 38648 selected companies belong to 63 distinct GICS industries and the occurrence of the different industries is quite heterogeneous. In the third and forth column of  Table~\ref{tabOrg} we list the occurrence of the 50 most common industries. The heterogeneity of the industries is immediately evident. In fact the most common industry  {\it Internet Software and Services} is characterizing 13891 companies whereas the {\it personal Products} industry (50th in rank) is characterizing  only 6 companies. The 18 industries with more than 100 occurrences  belongs to 7 out of 11 sectors. Specifically, we have 2 {\it Industrials} ( {\it Commercial Services \& Supplies} and  {\it Professional Services}), 2 {\it Consumer Discretionary} ( {\it Hotels, Restaurants \& Leisure} and  {\it Diversified Consumer Services}), 1 {\it Health Care} ({\it Health Care Technology}), 5 {\it Financials} ( {\it Capital Markets}, {\it Diversified Financial Services}, {\it Banks}, {\it Consumer Finance}, and {\it Insurance}), 5 {\it Information Technology} ( {\it Internet Software \& Services}, {\it Software}, {\it IT Services}, {\it Electronic Equipment, Instruments \& Components}, and {\it Technology Hardware, Storage \& Peripherals}), 2 {\it Communication Services} ( {\it Media} and  {\it Diversified Telecommunication Services}), and 1 {\it Real Estate} ({\it Real Estate Management \& Development}). Even when we limit to sizeable occurrences, the impact of the diffusion of fintech terms is on a broad number of economic sectors with a particular emphasis on Finance and Information technology. It is worth noting that the selected companies might be sometimes difficult to classify. In the above list of 18 top  industries, three of them are classified by connoting them as {\it ``Diversified"}. Moreover, the most frequent industry {\it Internet Software \& Services} is described by analysts as "a relatively small industry primarily engaged in enabling and supporting commerce and other types of business transactions over the Internet. So they offer cloud-based solutions and services that make customer interaction with businesses easier." \citep{web1} The definition of the industry within GICS was revised by Standard \& Poor's and MSCI/Barra companies \citep{web2} at the end of 2018. Reclassification events are occurring in several areas and carry information about technological evolution \citep{lafond2019long}. Here we interpret the reclassification event observed for the economic sector with the highest occurrence in the selected companies as an indication of the difficulty found by the analysts in defining nature and profile of the companies.

The third attribute we investigate is the municipality of the company location or headquarter. We have this information for 33,368 companies. They are located in 4474 distinct municipalities all over the world. The number of companies per municipality is again highly heterogeneous reflecting a Zipf like behavior \citep{Zipf1935,simon1960some}. In fact when we regress the logarithm of the number of companies on the logarithm of the rank of the municipality we obtain a power law exponent of $-1.073$ very close to the $-1$ value expected for a Zipf plot.

We observe a quite pronounced abundance of companies in some cities or metropolitan areas. The city with the largest number of companies is London UK. Other top cities are New York, San Francisco, and Singapore. In addition to San Francisco many other municipalities of the San Francisco Bay area are present in the top 50 municipalities (Palo Alto, San Jose, Mountain View, Menlo Park, San Mateo, Sunnyvale). By summing the number of companies operating in these municipalities of the San Francisco Bay area one obtains 2131 companies perhaps indicating the highest concentration of fintech companies in the world. Other metropolitan areas with a large number of companies are the great London area (1883 companies) and the New York City area (1738 companies). The list also contain small and medium size municipalities. One interesting example is the municipality of Zug in Switzerland having 116 companies (rank 46). The valley where this municipality of 120 thousand inhabitants is located is called the "crypto valley" and has hosted {\it The Crypto Valley Blockchain Conference} in 2019. On the other hand, the over-expression of companies with headquarters in Zug might also be related to the fact that Zug is a tax heaven for companies and the detected over-expression might only manifest the tendency of some of the companies dealing with fintech terms to locate their headquarters in a municipality with fiscal advantage.    

Heterogeneity, and most probably uneven coverage of companies across different countries, is therefore present for all three attributes. Our analysis will therefore use a methodology that is robust with respect to the presence of it. To properly deal with this heterogeneity we analyze relationships between company attributes and fintech terms as bipartite networks and we then detect over-expressed relationships. 

Specifically, we start our approach by constructing three bipartite networks. The first is a countries - fintech terms network where we aggregate all companies located in the same country, the second is a economic industries - fintech terms network where we aggregate all companies working in the same economic industry, and the third is a municipalities - fintech terms network where we aggregate all companies working in the same municipality. The first network is a bipartite network with 163 countries and 50 fintech terms. The number of links is 1651 and the link density is 0.203.  The second network is a bipartite network with 64 industries and 50 fintech terms. It has 707 links and a link density equals to 0.221.  The third network is a bipartite network with 4474 municipalities  and 50 fintech terms. In the third network links are 10893 and the link density is  0.048.

To highlight the over-expressed relationships between countries, industries, and municipalities with fintech terms we detect over-expressed links on all three networks. This is done by using the methodology of statistically validated network   \citep{tumminello2011statistically,hatzopoulos2015quantifying}. The detection of a statistically validated network (SVN) works as follows. Let us consider an attribute $a$ of companies whose occurrence is $N_a$ and a fintech term $b$ whose occurrence is $N_b$. Let us define $N_{a,b}$ as the number of occurrences of fintech term $b$ in documents of companies with attribute $a$ and let us call the total number of documents $N_t$.  With these definitions, the probability of observing $X$  co-occurrences of the attribute $a$ and fintech term $b$ under a null hypothesis of random mixing is well approximated by the hypergeometric distribution \citep{tumminello2011statistically}
\begin{equation}
\label{hypertheor}
H(X|N_t,N_a,N_b)=\frac{{N_a \choose X} {N_t-N_a \choose N_b-X}}{{N_t\choose N_b}}.
\end{equation}

The probability of Eq.~(\ref{hypertheor}) allows to estimate a p-value $p(N_{a,b})$ associated with the  empirical observation of $N_{a,b}$ co-occurrences or more of attribute $a$ and fintech term $b$. In fact, the  p-value is
\begin{equation}
\label{pvaltheor}
p(N_{a,b})=1-\sum_{X=0}^{N_{a,b}-1}H(X|N_t,N_a,N_b).
\end{equation}
With this approach, one can associate a p-value to all links of the bipartite network linking nodes of attributes of set $A$ and fintech terms of set $B$ by performing a statistical test. It is worth noting that the test highlights the over-expressions with respect to a null hypothesis that takes into account the heterogeneity of the attributes. In other words, the relationships highlighted by the test are not necessarily the most frequent but rather the ones that violates the null hypothesis assuming random connections between heterogeneous attributes and fintech terms.     

For each bipartite network, the number of statistical tests to perform is given by the number of links that are present in the bipartite network. This number is relatively high and for this reason a multiple hypothesis test correction is useful to avoid a large number of false positive. In the present investigation, we use the control of the false discovery rate (FDR) as multiple hypothesis test correction \citep{benjamini1995controlling} and we set to $0.01$ the value of the false discovery rate, i.e. the expected maximal fraction of false positive. 

We compute SVNs with a code written by us. However, programs computing SVNs from bipartite networks are available online \citep{bongiorno2017,challet2019}. Specifically, we have obtained SVNs of bipartite networks of (a) countries - fintech terms, (b) industries - fintech terms, and (c) municipalities - fintech terms.

\begin{figure}[h!]
\begin{center}
\includegraphics[width=1.0\linewidth]{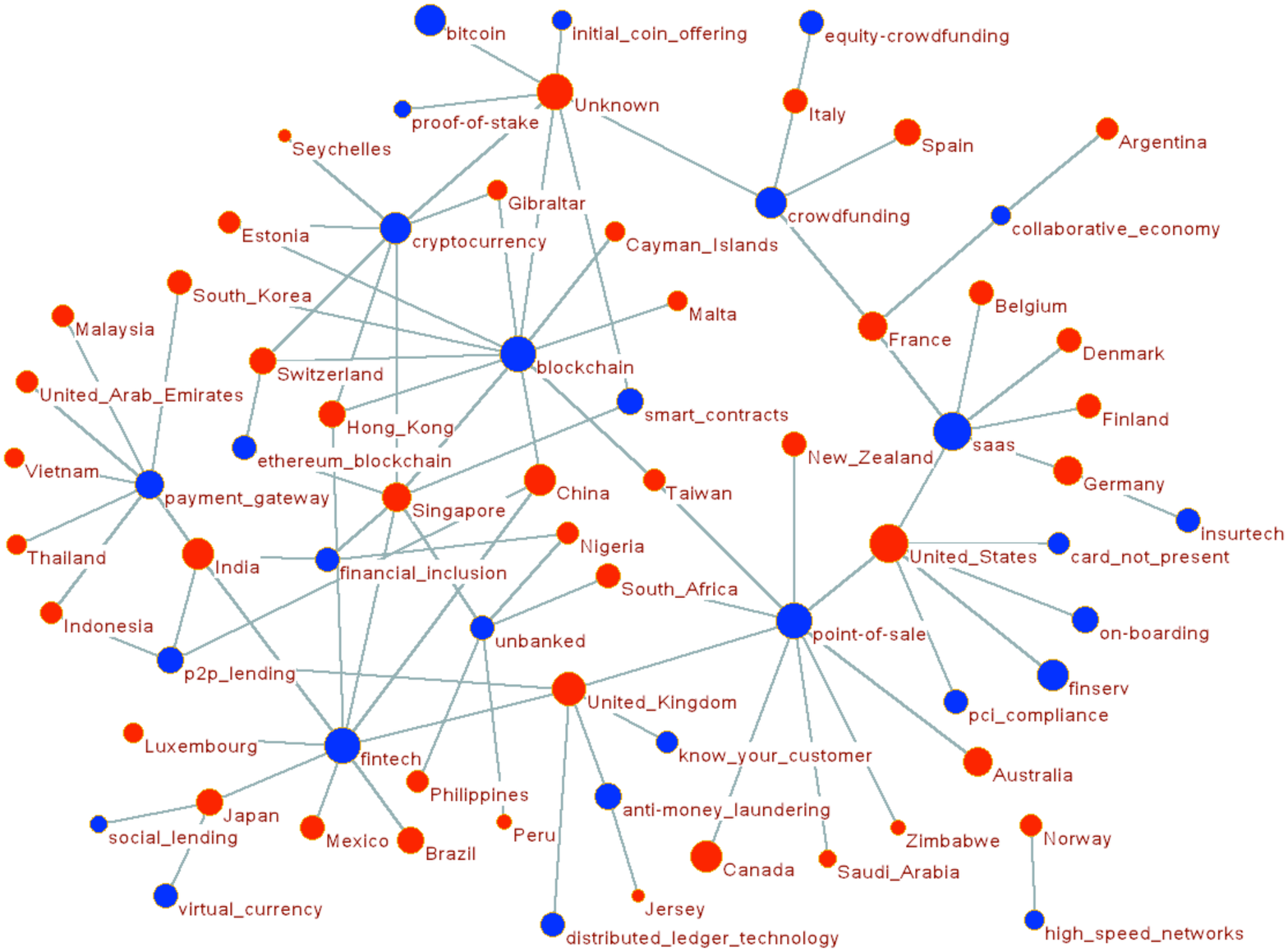}
\end{center}
\caption{Bipartite statistically validated network of countries - fintech terms. Blue nodes are fintech terms and red nodes are countries. For countries, the radius of each node is proportional to the logarithm of the number of companies of the country.  For fintech terms, the radius of each node is proportional to the logarithm of the term occurrence.}
\label{fig:B}
\end{figure}

The bipartite SVN of countries - fintech terms has 43 countries, 28 fintech terms, and 87 validated links. We are showing this network in Fig.~\ref{fig:B}. The blue nodes are fintech terms and the red nodes are countries.  All the companies not reporting the information about the country in the databases are labeled by the term ``Unknown''. In the figure the radius of each node describing a country (red nodes) is proportional to the logarithm of the number of companies of the country whereas the radius of each node describing a fintech term (blue nodes) is proportional to the logarithm of the term occurrence.

By analyzing the figure, we note that countries where companies present an over-expression of the word {\it Blockchain} in their profiles are  Gibraltar, Cayman Islands, Malta, Taiwan, China, Singapore, Hong Kong, Switzerland,  South Korea, and Estonia.  Mediterranean countries Italy, Spain and France have companies over-expressed in {\it Crowdfunding} whereas north European countries Belgium, Denmark, Finland and Germany present over-expression with {\it SAAS}. Germany has also an over-expressed link with {\it Insurtech}.  Fintech terms {\it Unbanked} and {\it Financial inclusion} are over-expressed in companies of the following countries: India, Singapore, Nigeria, South Africa, Peru and Philippines. All these countries except Singapore are developing countries with high potential of extension of financial inclusion.     

\begin{figure}[h!]
\begin{center}
\includegraphics[width=1.0\linewidth]{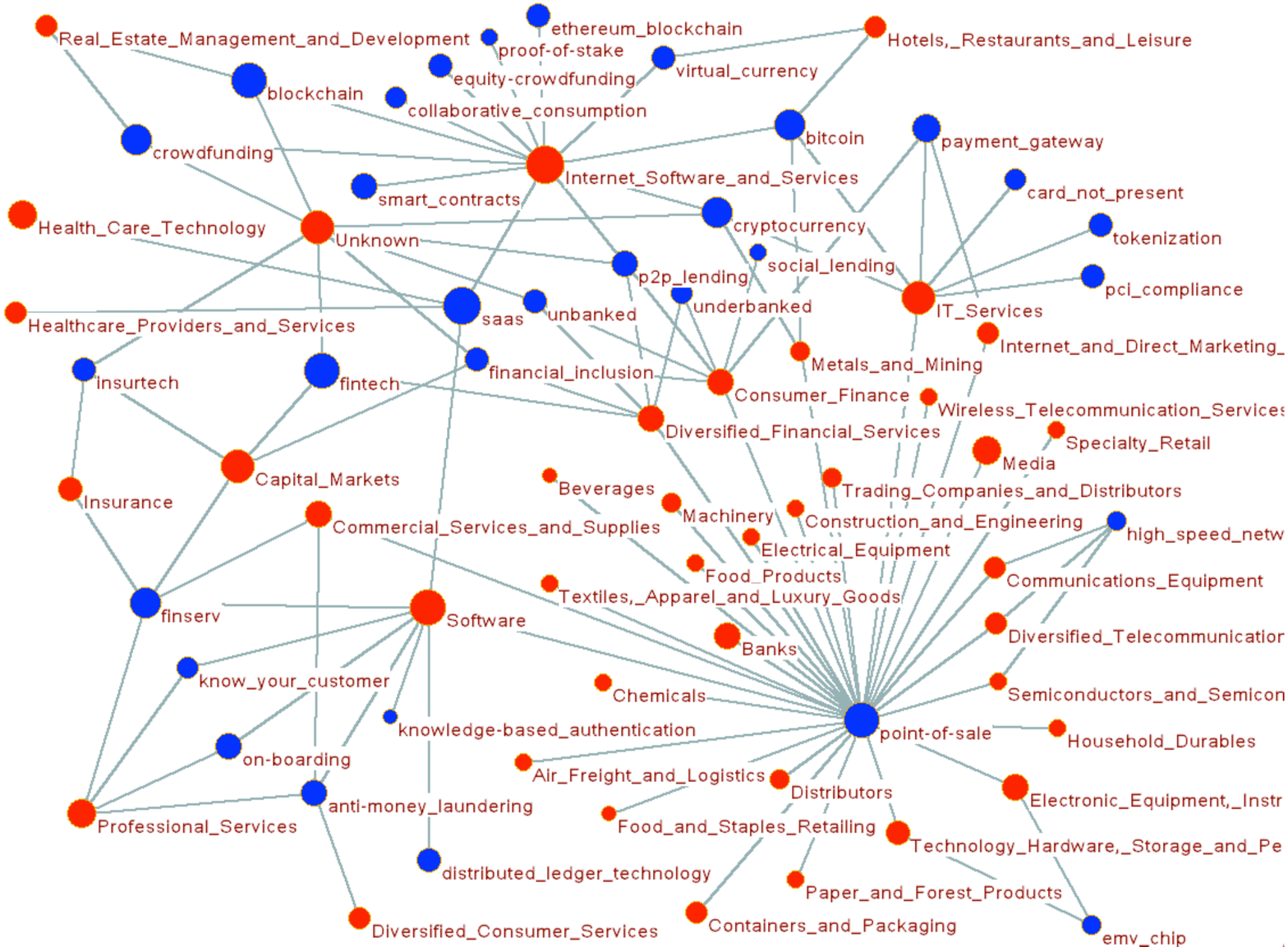}
\end{center}
\caption{Bipartite statistically validated network of industries - fintech terms. Blue nodes are fintech terms and red nodes are industries. For industries, the radius of each node is proportional to the logarithm of the number of companies of the industry.  For fintech terms, the radius of each node is proportional to the logarithm of the term occurrence.}
\label{fig:C}
\end{figure}

The bipartite SVN of industries - fintech terms has 40 industries, 31 fintech terms, and 101 validated links. The validated network is shown in Fig.~\ref{fig:C}. We note that the companies belonging to the {\it Internet Software \& Services} present  over-expression with some terms of the fintech list of terms. In fact the companies of this industry are linked with {\it Blockchain}, {\it Collaborative consumption}, {\it Equity crowdfunding}, {\it Proof of stake}, {\it Ethereum blockchain}, {\it Virtual currency}, {\it Bitcoin}, {\it Cryptocurrency}, {\it P2P lending}, {\it SAAS}, {\it Smart contract}, and {\it Crowdfunding}. Companies of the industry of {\it IT services} present over-expressed links with the fintech terms of  {\it Payment card industry (PCI) compliance}, {\it Tokenization}, {\it Card not present}, {\it Payment gateway}, {\it Bitcoin}, {\it Cryptocurrency}, and {\it Point of sale}. Companies belonging to the industry of {\it Software} or to the industry of {\it Professional services} present over-expressed links with {\it Finserv}, {\it Know your customer}, {\it On boarding}, and {\it Anti-money laundering}. Companies of the finance industries  {\it Capital markets}, {\it Diversified financial services}, and  {\it Consumer finance} are characterized by over-expression of the terms {\it Fintech}, {\it Finserv}, {\it Insurtech}, {\it Financial inclusion}, {\it Unbanked}, {\it Underbanked}, {\it P2P lending}, {\it Social lending}, {\it Payment gateway}, and {\it Point of sale}. It is also worth noting that several of the industries characterized by a limited number of companies (recognizable by nodes of small radius) are linked with {\it Point of sale}. Within fintech processes and services, this term is primarily used to address point of sale financing. Point of sales financing is the business practice allowing consumers to quickly finance large purchases with interest-free loans which are set up at the point of sale. Up until 2019, fintech firms have dominated this area.

\begin{figure}[h!]
\begin{center}
\includegraphics[width=1.0\linewidth]{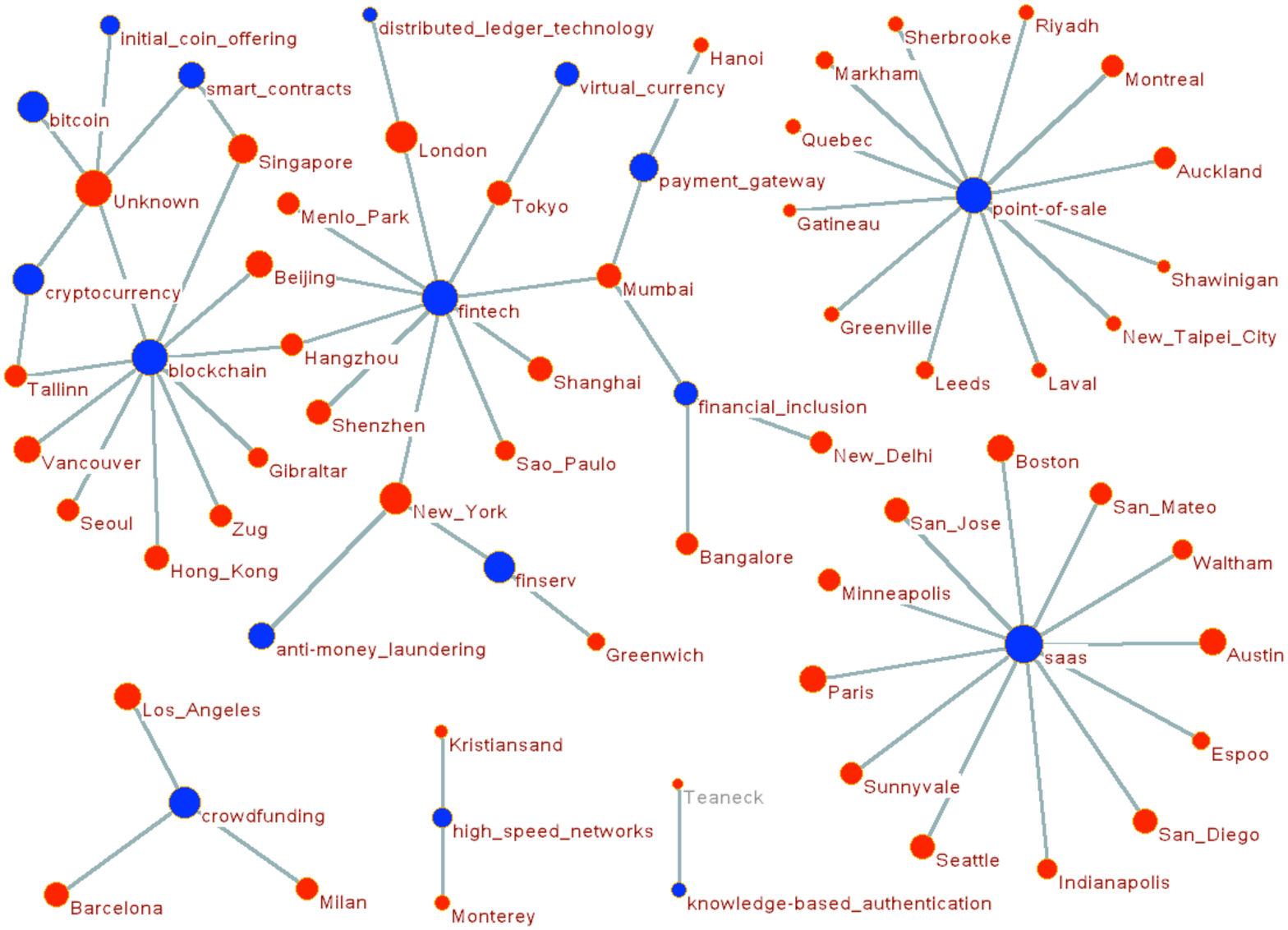}
\end{center}
\caption{Bipartite statistically validated network of municipalities - fintech terms. Blue nodes are fintech terms and red nodes are municipalities. For municipalities, the radius of each node is proportional to the logarithm of the number of companies of the municipality.  For fintech terms, the radius of each node is proportional to the logarithm of the term occurrence.}
\label{fig:D}
\end{figure}

The last bipartite SVN is the network of municipalities - fintech terms. The network detects 68 over-expressed links between 54 municipalities and 17 fintech terms. In Fig.~\ref{fig:D} we show the network. In this case the bipartite SVN shows several disjoint components. The largest component includes the fintech terms of {\it Cryptocurrency}, {\it Bitcoin}, {\it Initial coin offering}, {\it Smart contracts}, {\it Blockchain}, {\it Fintech}, {\it Distributed ledger technology}, {\it Virtual currency}, {\it Payment gateway}, {\it Financial inclusion}, {\it Finserv}, and {\it Anti-money laundering}.  It involves cities that are hosting the biggest financial centers of the world such as  {\it New York}, {\it Tokyo}, {\it Shanghai}, {\it Hong Kong}, {\it London},  {\it Shenzhen}, {\it Mumbai}, {\it Seoul}, and {\it Singapore}, and municipalities or cities with a strong tradition on digital innovation as {\it Menlo Park}, {\it Tallin}, and {\it Vancouver}. In the small municipality of Zug, companies present an over-expression of the term {\it Blockchain}, whereas the term {\it Financial inclusion} is over-expressed in companies located in {\it Mumbai}, {\it New Delhi}, and {\it Bangalore}.
The other components of the network are characterized by a single fintech term. Specifically, these fintech terms are {\it Software as a service (SAAS)},  {\it Point of sale},  {\it Crowdfunding},  {\it High speed networks}, and {\it Knowledge based authentication}. 

\section{Discussion and conclusions}
\label{Dco}
Our large scale textual analysis of news and blogs in English language shows that a set of terms has developed and consolidated during the calendar years from 2014 to 2018 ending up in a compact and coherent set of terms used worldwide to describe fintech business activities. The search for this set of terms in the professional descriptions of a large dataset  of companies located worldwide has faced the problem of the degree of coverage of databases in different countries. Databases are biased towards specific countries and therefore a simple frequency analysis can be misleading. We therefore perform an analysis using a network science approach that is able to detect over-expression of a specific attribute with respect to a null hypothesis taking into account the heterogeneity of the investigated bipartite network.

With our approach we obtain highlights about the over-expression of specific fintech terms in the description of a large number of companies of the fintech movement. Companies located both in developed and in developing economies present some degree of specialization (i.e. over-expression of occurrence of specific fintech terms in their professional description). Our analysis also shows that fintech topics, products and services have the potential to impact a large number of industries. In fact our analysis of the bipartite SVN economic sectors - fintech terms comprises 40 of the 63 economic sectors. One of the term with several statistically validated links, {\it point of sale}, is also used outside the field of fintech. We have retained this term in our analysis because it plays an important role in the fintech business. In fact, point of sale financing is one of the main areas of development of fintech activities. By considering the use of the term {\it point of sale} outside fintech, we acknowledge that some of its links might not be uniquely related to point of sale financing. However, it is worth noting that SVN approach is a pairwise approach and results obtained for a specific term do not affect results of other pairs. Therefore, in the unrealistic worst case that all links of  {\it point of sale} term do not relate to point of sale financing the remaining pairwise links between fintech terms and economic sectors would highlight over-expression of fintech terms in companies that are active in a minimum number of 22 distinct economic sectors.

We are also able to detect a geographical pattern of over-expression for companies dealing worldwide with fintech topics, services, and products. We characterize the geographical location down to the municipality of the headquarter of companies. The over-expressions detected show that, in addition to the most important financial centers, large number of companies are located in the San Francisco bay area and in a set of cities acting as innovation hubs of their countries. We are also able to highlight over-expression of small municipalities like Zug or Gibraltar that have clusters of companies with over-expression in the same area of the fintech business. Specifically, both municipalities have over-expression of blockchain in the descriptions of companies. 

In summary,  a methodology based on the analysis of bipartite networks constructed from biased or incomplete databases is able to highlight over-expressions of attributes of elements of the systems (in the present case companies). Our methodology is characterized by the control of false positives in the determination of statistically significant over-expressions. In other words, the over-expressions detected are all statistically significant at the chosen level of the control of false discovery rate ($\alpha=0.01$). Unfortunately, a methodology simultaneously controlling the number of false positives and the number of false negatives is not yet available and therefore we cannot exclude a sizeable number of false negatives.  

In spite of this limitation, by relying on a full control of absence of false positives our analysis unequivocally shows that fintech is a multi industry, geographically distributed movement with a detectable level of geographical and economic sector specialization. This business movement is focusing on technical and methodological innovation of financial products, services, and activities. The innovations produced have the potential to deeply change the way mankind is dealing with finance in the coming years.\\

{\bf{Conflict of Interest Statement}}

Author F.C. is employed by company Quid, San Francisco, USA. R.N.M. declares no competing interests.\\ 

{\bf{Author Contributions}}

F.C. and R.N.M. conceived the study. F.C. performed the text analysis of databases. F.C. and R.N.M. analyzed and interpreted the results and wrote the manuscript.\\

{\bf{Data Availability Statement}}

The data that support the findings of this study are available from Capital IQ, Crunchbase and LexisNexis. Restrictions apply to the availability of these data, which were used under license for this study. Requests to access these datasets should be directed to  Crunchbase 
\url{https://about.crunchbase.com/products/crunchbase-pro/},
LexisNexis 
\url{https://www.lexisnexis.com/en-us/products/nexis/feature-get-the-story.page},
and S\&P Global (for Capital IQ)
\url{https://www.spglobal.com/marketintelligence/en/solutions/sp-capital-iq-platform} .%


%
%

%

\begin{acknowledgments}
We thank Luca Marotta for his help in preparing the alluvial diagram. RNM acknowledge financial support of the project ``Stochastic forecasting in complex systems", project code: 2017WZFTZP.
\end{acknowledgments}

\bibliography{Chaos_CM_R}

\end{document}